\documentclass{PoS}

\usepackage{amsmath}

\newcommand{\dedx}{\langle dE/dx\rangle}
\newcommand{\dNdeta}{\langle \mathrm{d}N_{\mathrm{ch}}/\mathrm{d}\eta\rangle}

\title{Measurements of the fluctuations of identified particles in ALICE at the LHC}

\ShortTitle{Measurements of the fluctuations of identified particles in ALICE at the LHC}

\author{\speaker{Alice Ohlson} for the ALICE Collaboration\\
       Ruprecht-Karls-Universität Heidelberg\\
       E-mail: \email{alice.ohlson@cern.ch}}

\abstract{The event-by-event fluctuations of identified particles in ultrarelativistic nucleus-nucleus collisions give information about the state of matter created in these collisions as well as the phase diagram of nuclear matter.  In this proceedings, we present the latest results from ALICE on the centrality and pseudorapidity dependence of net-proton fluctuations, which are closely related to net-baryon fluctuations, as well as net-kaon and net-pion fluctuations.  The effects of volume fluctuations and global baryon conservation on these observables are discussed.  Furthermore, the correlated fluctuations between different particle species, quantified by the observable $\nu_{dyn}$, are also shown as functions of multiplicity and collision energy and are compared with Monte Carlo models.  These measurements are performed in Pb--Pb collisions at $\sqrt{s_{\mathrm{NN}}} = 2.76$ TeV using the novel Identity Method and take advantage of the excellent particle identification capabilities of ALICE.}

\FullConference{Critical Point and Onset of Deconfinement\\
         7-11 August, 2017\\
         The Wang Center, Stony Brook University, Stony Brook, NY}

\begin{document}

\section{Introduction}

Ultrarelativistic collisions of heavy nuclei create a hot, dense, strongly-interacting state of matter in which quarks and gluons are deconfined.  Studying the properties of this state, known as the quark-gluon plasma (QGP), allows us to understand the high-temperature and high-density regime of the phase diagram of nuclear matter.  Measurements of event-by-event fluctuations of particle multiplicities probe the properties and phase structure of strongly-interacting matter.  These quantities are of particular interest because they are related, under a set of assumptions, to the thermodynamic susceptibilities of the medium.  

The thermodynamic susceptibilities, $\chi$, are a set of observables which characterize the properties of a thermodynamic system by describing its response to changes in external conditions.  Of particular interest in this case are the $n^{th}$-order derivatives of the reduced pressure ($P/T^4$) with respect to the reduced chemical potential ($\mu_N/T$), 
\begin{equation}
\chi_n^{N=B,S,Q} = \frac{\partial^n (P/T^4)}{\partial (\mu_N/T)^n},
\end{equation}
where $N$ denotes an additive quantum number such as baryon number ($B$), strangeness ($S$), or electric charge ($Q$) and the corresponding chemical potentials are $\mu_B$, $\mu_S$, and $\mu_Q$, respectively.  

The susceptibilities $\chi_n^{B,S,Q}$ can be calculated in lattice quantum chromodynamics (lattice QCD, or LQCD) within the grand canonical ensemble (GCE) where they are related to the central moments of the multiplicity distribution of conserved charges.  The relationships between the higher moments and the susceptibilities $\chi^{B,S,Q}_n$ can be given by
\begin{equation}
\begin{split}
&M = \langle \Delta N\rangle = VT^3\chi_1,\\
&\sigma^2 = \langle (\Delta N-\langle \Delta N\rangle)^2\rangle = VT^3\chi_2,\\
&S = \langle (\Delta N-\langle \Delta N\rangle)^3\rangle/\sigma^3 = \frac{VT^3\chi_3}{(VT^3\chi_2)^{3/2}},\\
&\kappa = \langle (\Delta N-\langle \Delta N\rangle)^4\rangle/\sigma^4 - 3 = \frac{VT^3\chi_4}{(VT^3\chi_2)^{2}}.
\end{split}
\label{eq:moments}
\end{equation}
where $\Delta N$ is the net number of charges ($\Delta N_{B,S,Q} = N_{B,S,Q}-N_{\bar{B},\bar{S},\bar{Q}}$), $V$ is the volume of the system, and $T$ is the temperature.  If the volume and temperature are constant, then the factors of $VT^3$ can be eliminated by measuring products of the moments: 
\begin{equation}
\begin{split}
&S\sigma = \chi_3/\chi_2\\
&\kappa\sigma^2 = \chi_4/\chi_2.
\end{split}
\label{eq:ratios}
\end{equation}

While LQCD calculations become difficult where $\mu_B$ is non-zero, high-energy heavy-ion collisions at the Large Hadron Collider (LHC) probe the region of the phase diagram very close to $\mu_B = 0$, and therefore studying event-by-event net-baryon number fluctuations in these collisions allows us to test precise LQCD predictions.  Additionally, these measurements also look for signs of criticality which may persist even far from the phase transition.  

However, there are multiple effects which cause the relationship between the theoretically-calculable quantities and the experimentally-measurable observables to be inexact.  First, the \\experimentally-accessible quantities are the number of particles of a given species, not the strangeness or baryon number.  Typically, unidentified charged particles ($\Delta N_Q = N_+ - N_-$) or charged pions ($\Delta N_{\pi} = N_{\pi^+} - N_{\pi^-}$) are used as proxies to study net-charge fluctuations, charged kaons ($\Delta N_{K} = N_{K^+} - N_{K^-}$) are measured to access net-strangeness fluctuations, and net-proton ($\Delta N_{p} = N_{p} - N_{\bar{p}}$) moments are used as a proxy for baryon number fluctuations.  Second, while in LQCD the volume of the system can be fixed thus allowing the $V$ terms to cancel in Eq.~\ref{eq:ratios}, the system volume is not experimentally accessible and therefore volume fluctuations are intrinsic to the measurement.  Furthermore, when measurements are done within a fixed pseudorapidity range ($\Delta\eta$) the system can be viewed as sitting within a particle bath, but when $\Delta\eta$ is large global conservation laws cause the GCE approximation to break down.  Each of these effects should be explored and taken into account when interpreting experimental measurements and their comparison to theoretical calculations.  

\section{Experimental setup \& Analysis technique}

The measurements presented here were performed in Pb--Pb collisions at a center-of-mass energy per nucleon-nucleon pair of $\sqrt{s_{\mathrm{NN}}} = 2.76$~TeV at the LHC using the ALICE (A Large Ion Collider Experiment) detector.  The principle subsystems used in the following measurements were the ITS (Inner Tracking System) and TPC (Time Projection Chamber) for charged particle tracking, and the V0 detectors in the forward region ($-3.7 < \eta < -1.7$ and $2.8 < \eta < 5.1$) for event centrality determination.  Particle identification was performed in the TPC from the specific energy loss ($\dedx$) of charged tracks.  For more details on the ALICE experiment, see Ref.~\cite{ALICEdet}.  

The results presented below were obtained using the Identity Method~\cite{identity1,identity2,identity3}, which makes it possible to calculate the moments of the identified particle multiplicity distribution even when particle identification is done on a statistical, i.e. not track-by-track, basis.  First, the full, inclusive $\dedx$ distribution is obtained from a large sample of events. This allows the probability that a given track corresponds to a particular particle species to be determined with high precision.  Each track is assigned a weight $w_{\pi,K,p,e}$ between 0 and 1, which corresponds to the probability that a particle is a pion, kaon, proton, or electron.   The sum of all track weights in a particular event, $W_{\pi,K,p,e} = \sum w_{\pi,K,p,e}$, is then calculated, and the distributions of $W_{\pi,K,p,e}$ are obtained.  The Identity Method then provides a mathematical formalism for unfolding the moments of the $W_{\pi,K,p,e}$ distributions into the moments $\langle N_{\pi,K,p,e}^n\rangle$.  

Traditional particle identification techniques reduce contamination by using additional detector information or rejecting altogether particles which fall in regions of phase space where the identification is unclear, thus lowering the detection efficiency of the particles of interest.  On the other hand, the Identity Method explicitly accounts for the effects of imprecise particle identification without lowering the detection efficiency.  

\section{Net-proton, net-kaon, and net-pion fluctuations}

\begin{figure}[tb]
\centering
\includegraphics[width=0.45\linewidth]{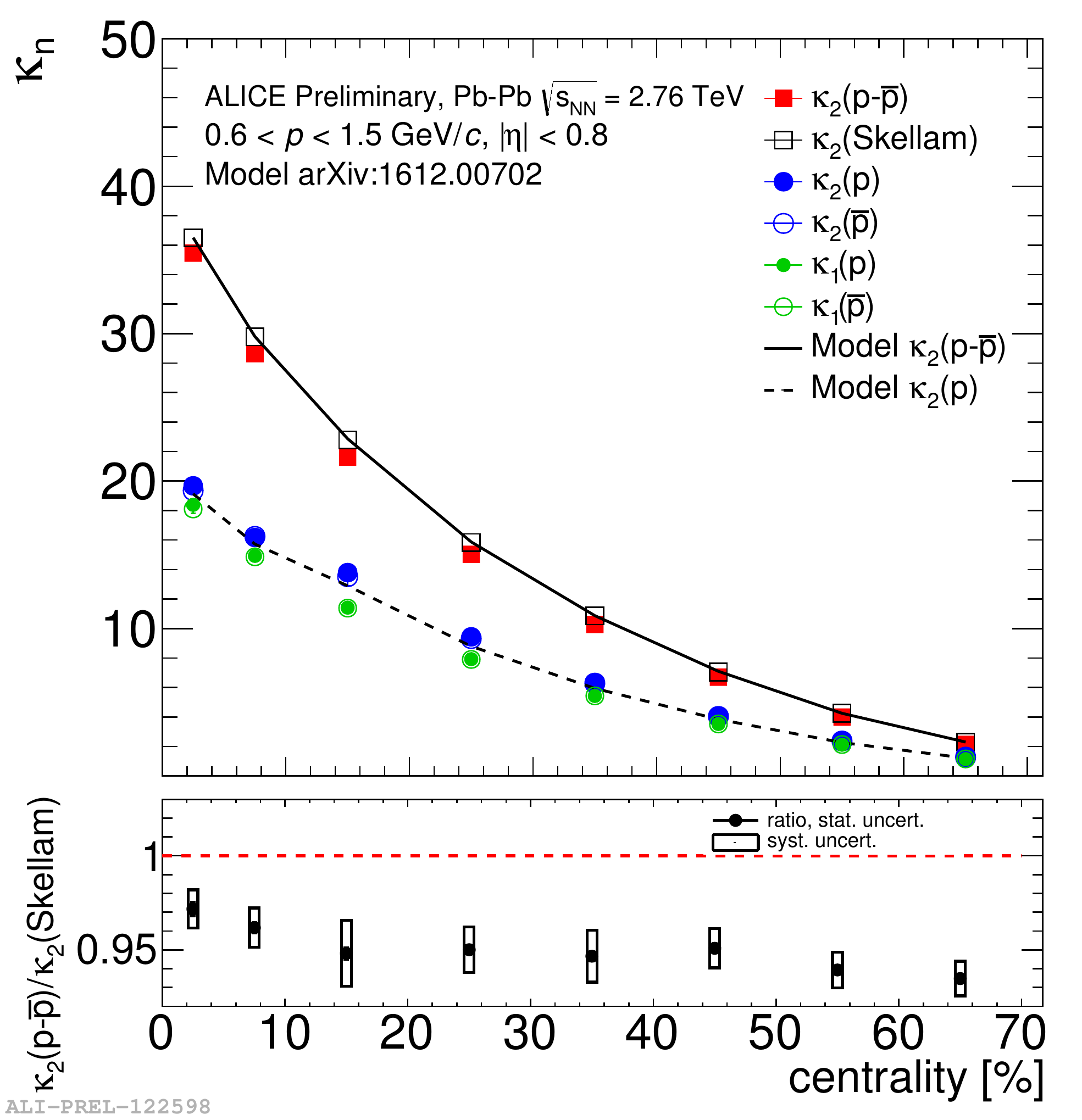}~~
\includegraphics[width=0.45\linewidth]{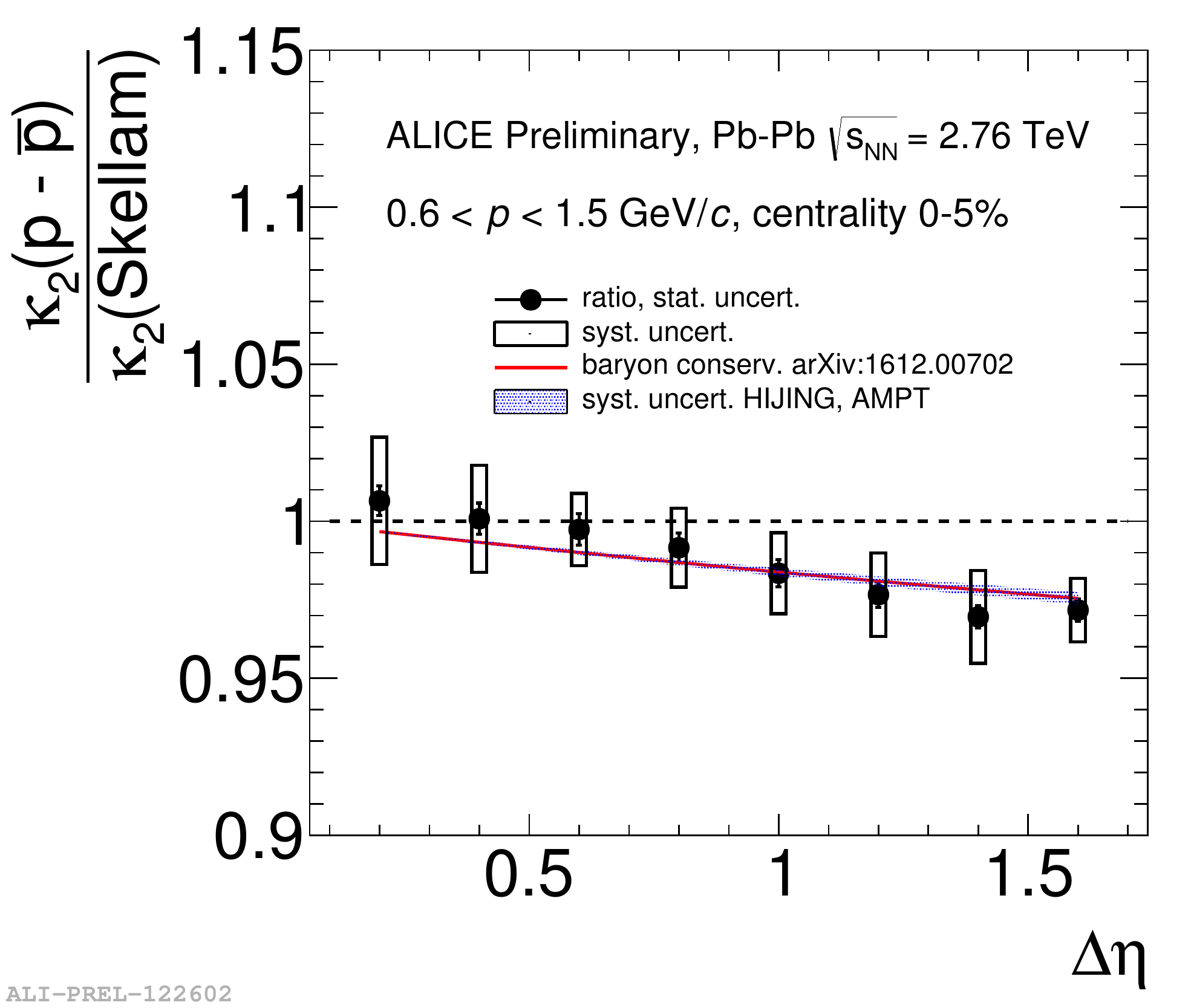}
\caption{(left) The first ($\kappa_1$) and second ($\kappa_2$) moments of protons ($p$), antiprotons ($\bar{p}$), and net-protons ($p-\bar{p}$) are measured as a function of centrality and compared with the Skellam expectation ($\kappa_2(Skellam) = \kappa_1(p)+\kappa_1(\bar{p})$).  (right) The second moment of net-protons, compared to the Skellam expectation, is measured as a function of the pseudorapidity acceptance of the measurement ($\Delta\eta$).  The results in both figures are compared with a model~\cite{netbaryon} which includes the effects of volume fluctuations due to the experimental centrality determination procedure as well as global baryon number conservation.\label{fig:netprotons}}
\end{figure}

Figure~\ref{fig:netprotons} shows the measurement of the first and second moments of the proton ($\kappa_{n=1,2}(p)$) and antiproton ($\kappa_{n=1,2}(\bar{p})$) multiplicity distributions: 
\begin{align}
\kappa_1(p) = \langle N_p\rangle, ~~~&\kappa_1(\bar{p}) = \langle N_{\bar{p}} \rangle\\
\kappa_2(p) = \langle \left(N_p - \langle N_p\rangle\right)^2\rangle, ~~~&\kappa_2(\bar{p}) = \langle \left(N_{\bar{p}} - \langle N_{\bar{p}}\rangle\right)^2\rangle
\end{align}
The measured second moment of the net-proton multiplicity ($\Delta N_p = N_p - N_{\bar{p}}$) distribution, defined in Eq.~\ref{eq:netproton}, is also shown as a function of event centrality in Fig.~\ref{fig:netprotons}.  
\begin{align}
\label{eq:netproton}\kappa_2(p-\bar{p}) &= \langle \left(\Delta N_p - \langle \Delta N_p\rangle\right)^2\rangle\\
&= \langle \left(N_p - N_{\bar{p}} - \langle N_p - N_{\bar{p}}\rangle\right)^2\rangle\\
&= \kappa_2(p) + \kappa_2(\bar{p}) - 2\left(\langle N_p N_{\bar{p}}\rangle - \langle N_p\rangle\langle N_{\bar{p}}\rangle\right)
\end{align}
If the multiplicity distributions of $N_p$ and $N_{\bar{p}}$ are Poissonian and uncorrelated, then the distribution of $\Delta N_p$ is Skellam.  The higher moments of a Skellam distribution are simply related to the first moments of the individual particles, in particular $\kappa_2(Skellam) = \kappa_1(p)+\kappa_1(\bar{p})$.  In Fig.~\ref{fig:netprotons} a deviation from the Skellam baseline is observed.  However, a model~\cite{netbaryon} which includes the effects of participant fluctuations on the experimental results shows good agreement with the data.  The model takes as input only the mean multiplicities, $\kappa_1(p)$ and $\kappa_1(\bar{p})$, and the experimental centrality determination procedure, and reproduces $\kappa_2(p)$, $\kappa_2(\bar{p})$, and $\kappa_2(p-\bar{p})$ within a consistent framework without the need of correlations or critical fluctuations.  

Furthermore, the same model (\cite{netbaryon}) incorporates the effects of global baryon conservation which can be observed in the dependence of the second moments on the pseudorapidity acceptance of the measurement ($\Delta\eta$), also shown in Fig.~\ref{fig:netprotons}.  Within the model, the deviation from a Skellam distribution can be parameterized by Eq.~\ref{eq:model}, where $\langle N_p^{meas}\rangle$ is the number of protons within the acceptance range $\Delta\eta$ and $\langle N_B^{4\pi}\rangle$ is the total number of baryons in the full $4\pi$ phase-space.  
\begin{equation}
\frac{\kappa_2(p-\bar{p})}{\kappa_2(Skellam)} = 1-\frac{\langle N_p^{meas}\rangle}{\langle N_B^{4\pi}\rangle}
\label{eq:model}
\end{equation}
The factor $\langle N_B^{4\pi}\rangle$ is obtained by extrapolating from the number of baryons within the acceptance, $\langle N_B^{acc} \rangle$, determined in Ref.~\cite{spectra}.  The Monte Carlo generators HIJING and AMPT are used for the extrapolation, and the small differences between the two generators are included in the systematic uncertainties on the model.  As was observed in the centrality dependence of $\kappa_2(p-\bar{p})$, the model can also fully describe the $\Delta\eta$ dependence as due to baryon conservation without the need of critical fluctuations.  

\begin{figure}[tb]
\centering
\includegraphics[width=0.32\linewidth]{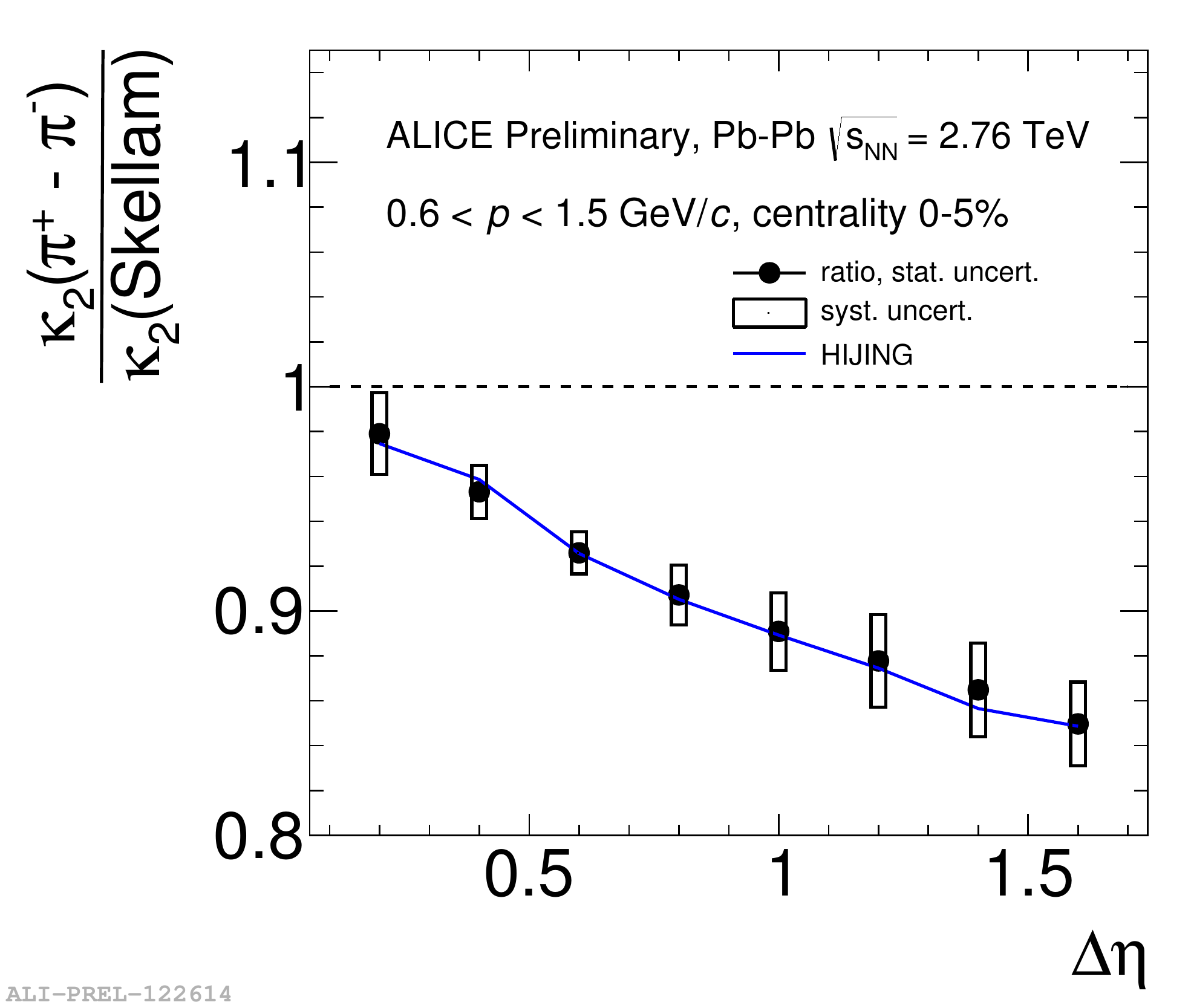}
\includegraphics[width=0.32\linewidth]{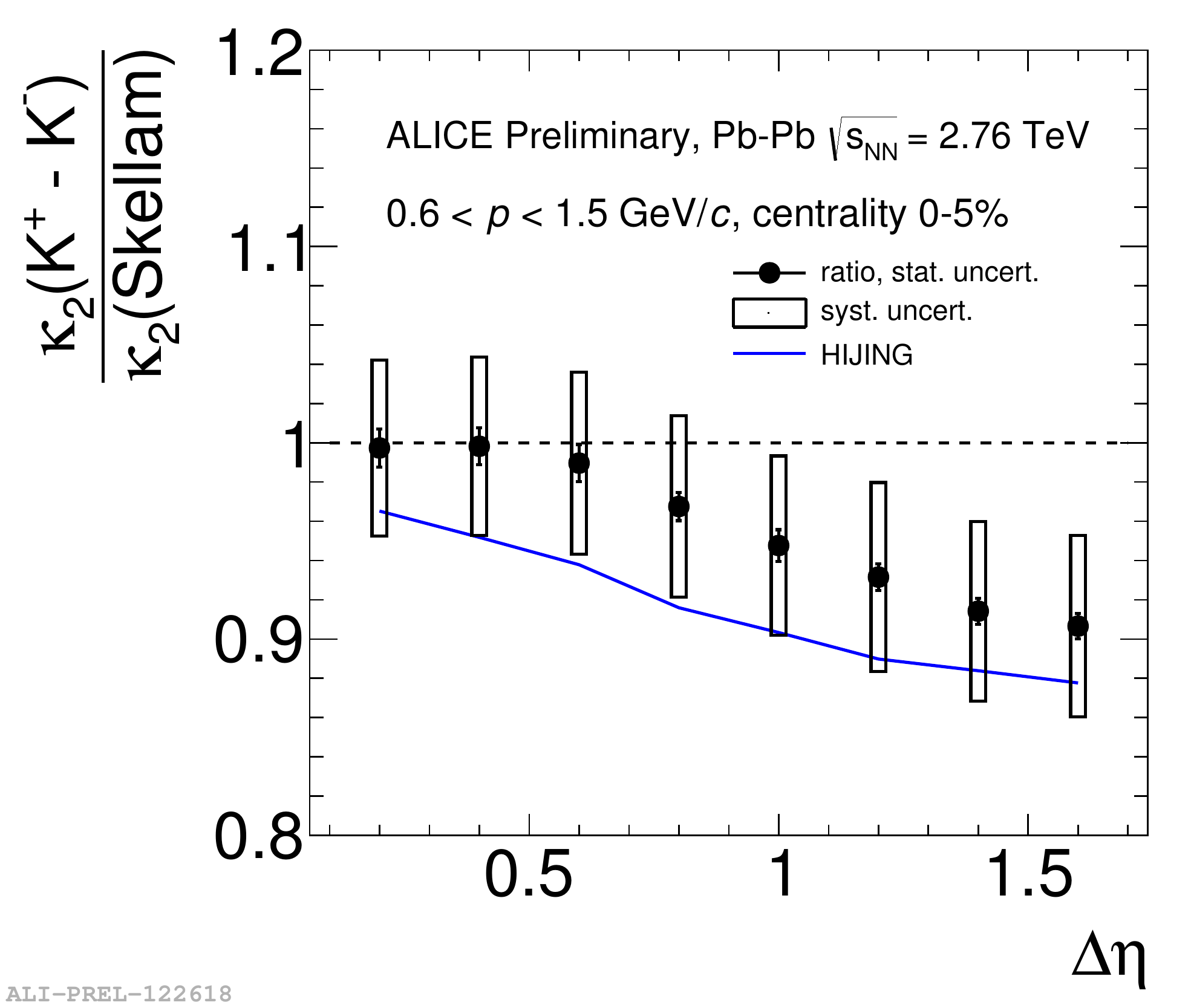}
\includegraphics[width=0.32\linewidth]{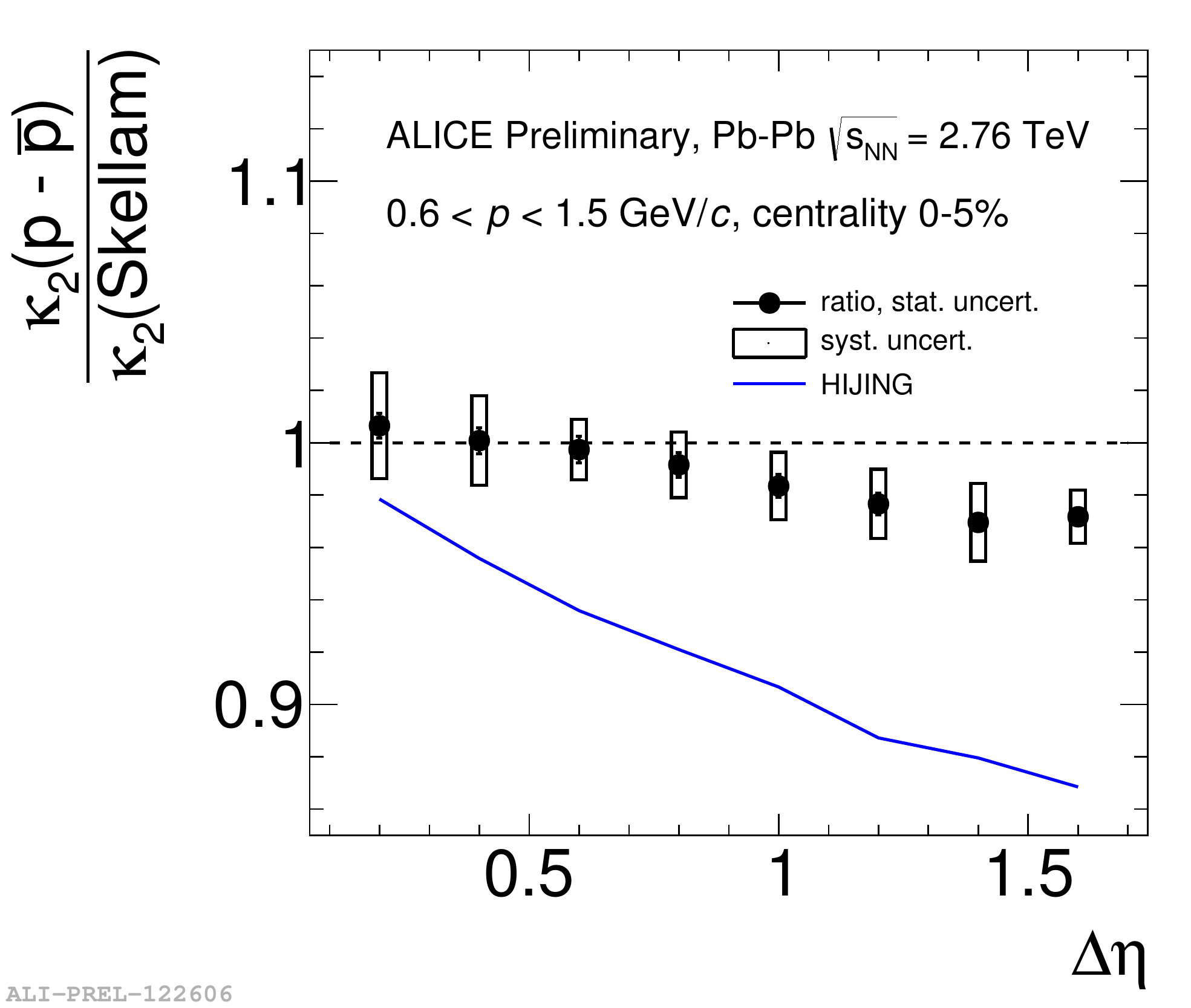}
\caption{The $\Delta\eta$ dependence of the (left) net-pion, (center) net-kaon, (right) net-proton second moments in 0-5\% central Pb--Pb collisions are compared to predictions from the HIJING Monte Carlo generator.\label{fig:netparticle}}
\end{figure}

In Fig.~\ref{fig:netparticle}, the results on the centrality dependence of the second moments of net-pion, net-kaon, and net-proton distributions are compared with HIJING.  While the pions show good agreement with HIJING, the agreement with the kaons is marginal, and the protons show significant disagreement.  However, the production of pions and kaons from resonance decays contributes significantly to the measured $\kappa_2(\pi^+-\pi^-)$ and $\kappa_2(K^+-K^-)$, which also means that a Skellam distribution is not the proper baseline for these measurements.  

\section{Identified particle fluctuations}

In addition to measuring the fluctuations of particles and their antiparticles, the event-by-event correlated fluctuations of different species are also investigated.  The second moment of the relative abundances between two particle types, $A$ and $B$, can be written with the variable $\nu$, defined in Eq.~\ref{eq:nu}.  
\begin{align}
\label{eq:nu}\nu &= \left\langle \left( \frac{N_A}{\langle N_A\rangle} -  \frac{N_B}{\langle N_B\rangle} \right)^2 \right\rangle\\
&= \frac{\langle N_A^2 \rangle}{\langle N_A\rangle^2} + \frac{\langle N_B^2 \rangle}{\langle N_B\rangle^2} - 2\frac{\langle N_A N_B\rangle}{\langle N_A\rangle \langle N_B\rangle}\label{eq:nu2}
\end{align}
In Eq.~\ref{eq:nu3} the independent statistical fluctuations of $N_A$ and $N_B$ are subtracted from $\nu$ to obtain a measure of the dynamical fluctuations, called $\nu_{dyn}$, defined in Eq.~\ref{eq:nudyn}.
\begin{align}
\label{eq:nu3}\nu_{dyn} &= \nu - \left(\frac{1}{\langle N_A\rangle} + \frac{1}{\langle N_B\rangle}\right)\\
&= \frac{\langle N_A\left( N_A -1\right) \rangle}{\langle N_A\rangle^2} + \frac{\langle N_B \left( N_B - 1\right) \rangle}{\langle N_B\rangle^2} - 2\frac{\langle N_A N_B\rangle}{\langle N_A\rangle \langle N_B\rangle}
\label{eq:nudyn}
\end{align}
If $N_A$ and $N_B$ have Poisson distributions and are uncorrelated, then $\nu_{dyn} = 0$.   

\begin{figure}[tb]
\centering
\includegraphics[width=0.49\linewidth]{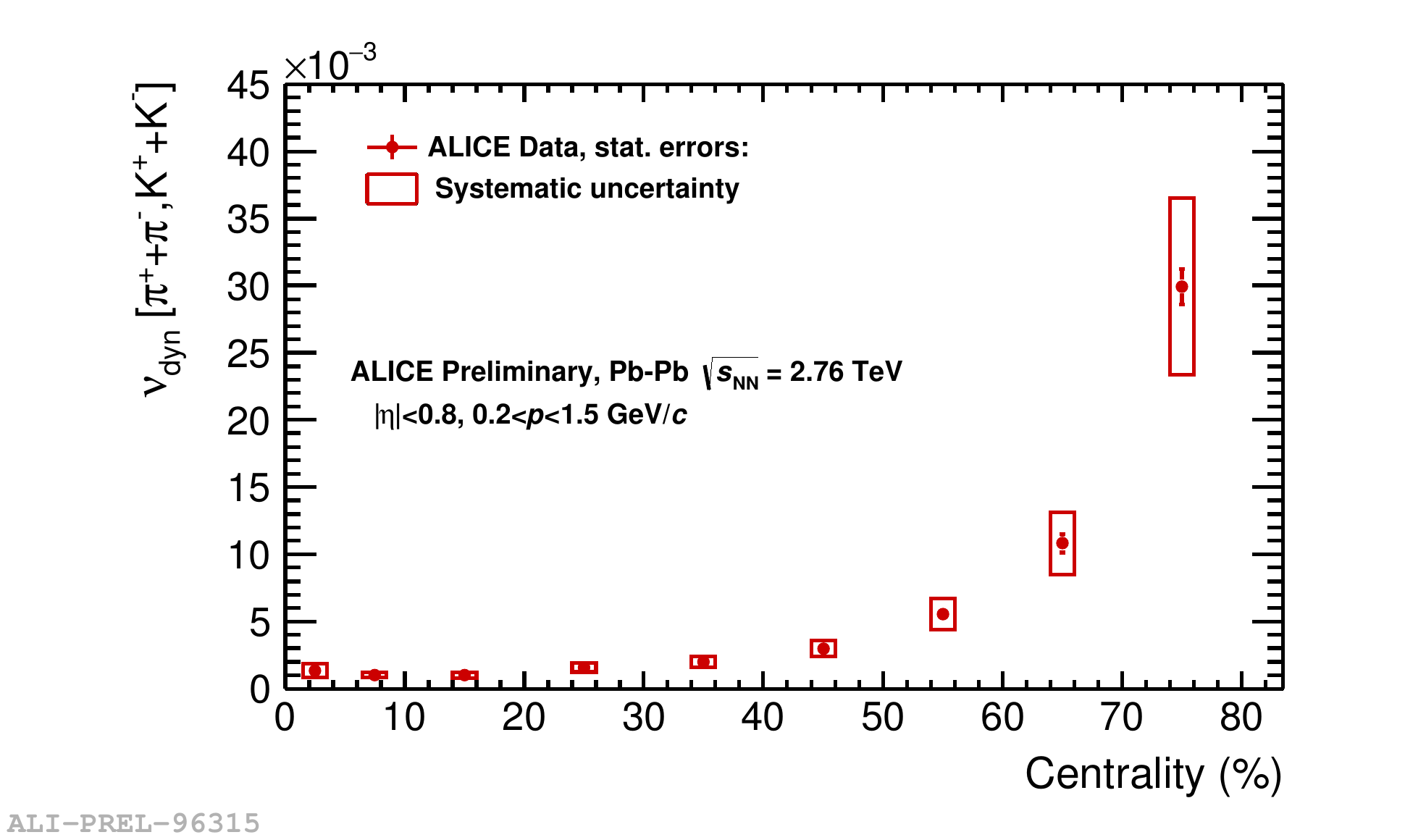}
\includegraphics[width=0.49\linewidth]{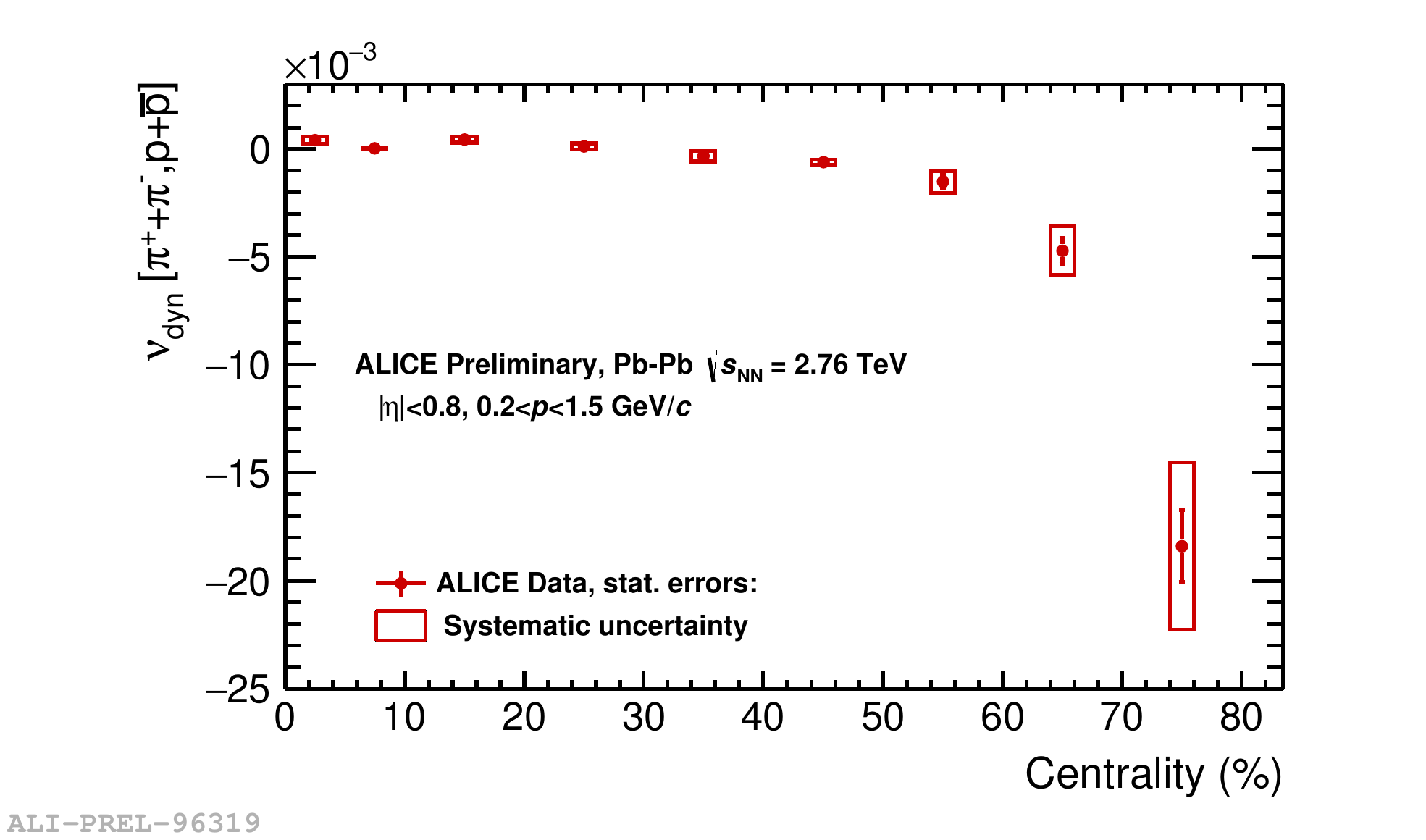}\\
\includegraphics[width=0.49\linewidth]{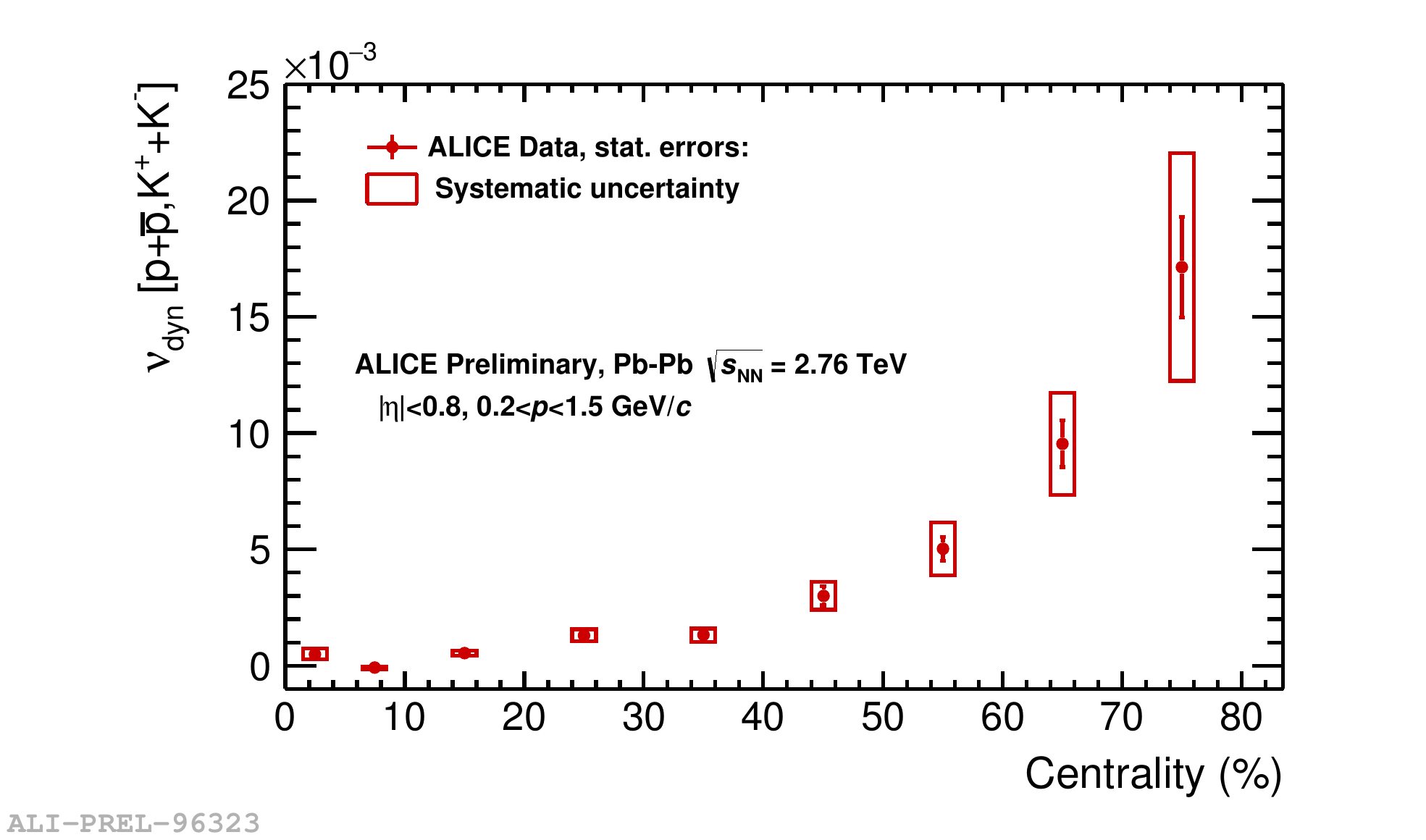}
\caption{The centrality dependence of (top left) $\nu_{dyn}[\pi,K]$, (top right) $\nu_{dyn}[\pi,p]$, and (bottom) $\nu_{dyn}[p,K]$ is measured.\label{fig:nudyn}}
\end{figure}

\begin{figure}[tb]
\centering
\includegraphics[width=0.45\linewidth]{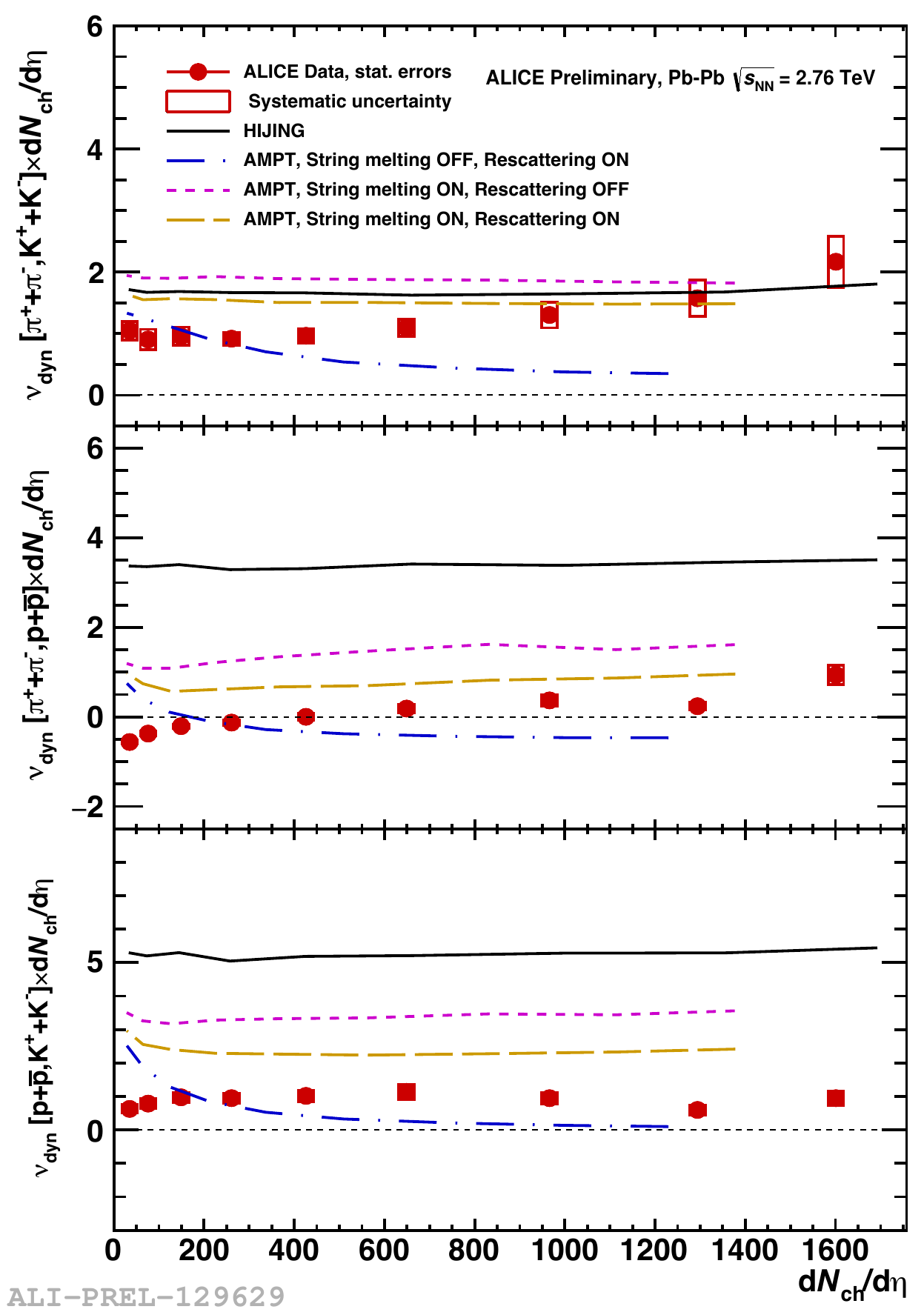}
\includegraphics[width=0.5\linewidth]{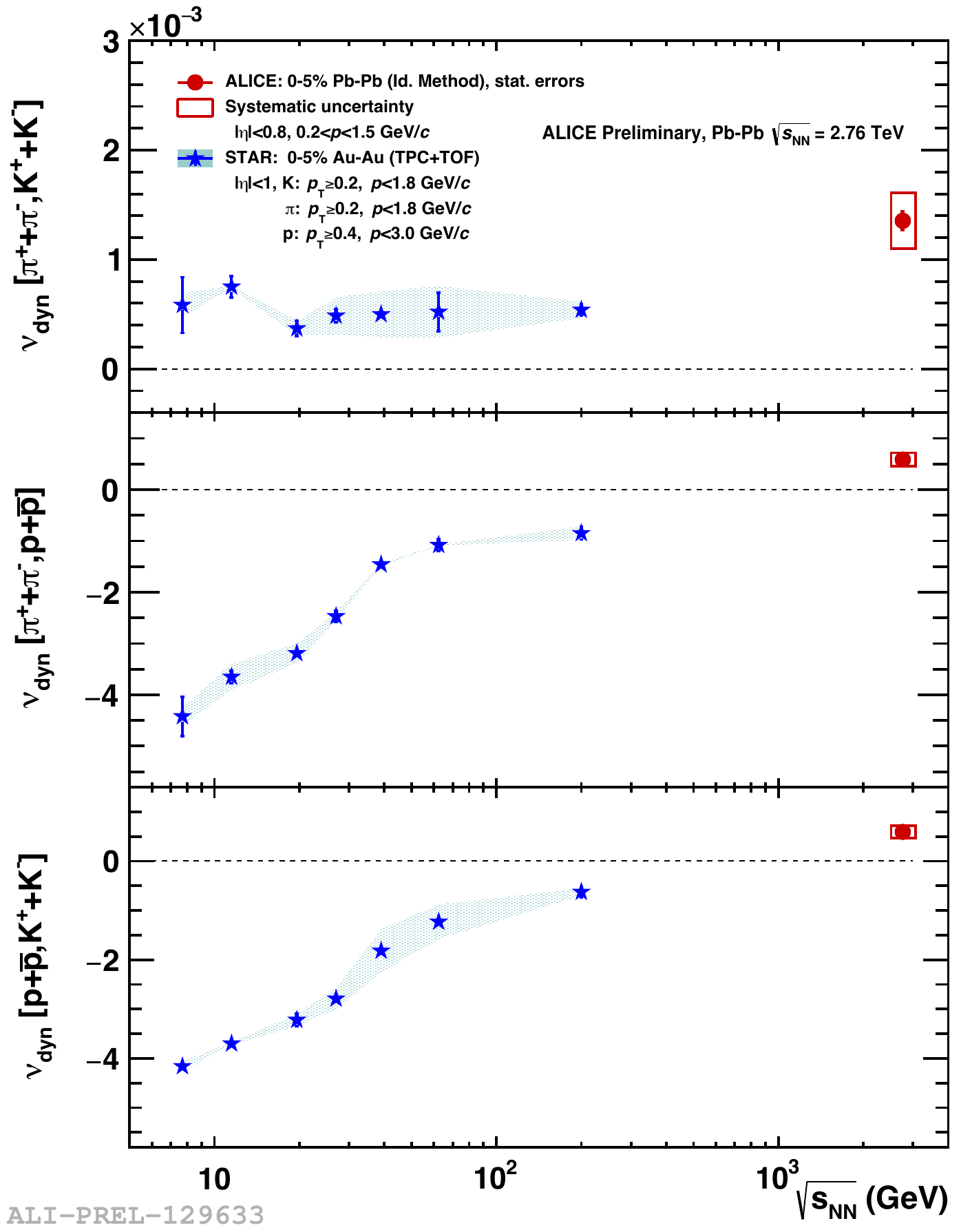}\\
\caption{(left) The measurements of $\nu_{dyn}[\pi,K]$, $\nu_{dyn}[\pi,p]$, and $\nu_{dyn}[p,K]$ are scaled by the multiplicity $\dNdeta$ and compared with HIJING and three configurations of AMPT.  (right) The values of $\nu_{dyn}$ measured in 0-5\% Pb--Pb and Au--Au collisions are shown as a function of center-of-mass energy.\label{fig:nudyn2}}
\end{figure}

Figure~\ref{fig:nudyn} shows the results for the correlated fluctuations of each combination of pions ($\pi^++\pi^-$), kaons ($K^++K^-$), and protons ($p+\bar{p}$): $\nu_{dyn}[\pi,K]$, $\nu_{dyn}[p,K]$, and $\nu_{dyn}[\pi,p]$.  The values of $\nu_{dyn}$ are small in central events but grow in peripheral events.  This deviation from the Poisson expectation may have contributions from other sources of correlations such as jets and resonance decays, but is also due to an intrinsic multiplicity scaling in the $\nu_{dyn}$ observable.  This multiplicity scaling is removed by multiplying by $\dNdeta$, as shown in Fig.~\ref{fig:nudyn2}, where it is observed that the data still deviates from the Poisson baseline.  Comparisons with the AMPT and HIJING model show qualitative agreement with the data, but quantitative differences.  

Finally, the ALICE results in the 0-5\% centrality range at a center-of-mass energy of $\sqrt{s_{\mathrm{NN}}} = 2.76$~TeV are compared with results from the STAR beam energy scan in Fig.~\ref{fig:nudyn2}.  While a sign change of $\nu_{dyn}[p,K]$, and $\nu_{dyn}[\pi,p]$ is observed, the evolution from RHIC to LHC energies is generally smooth.  

\section{Conclusions}

In this proceedings the latest results from ALICE on the fluctuations of identified particles in Pb--Pb collisions at the LHC are presented.  The measurements exploit the excellent particle identification capabilities of the ALICE detector, and are performed with the Identity Method in order to account for cases in which the particle identification is unclear without lowering the detection efficiency.  

The fluctuations of net-protons, net-kaons, and net-pions are of particular interest because they are related to the susceptibilities of the QGP matter which can be calculated within LQCD.  The centrality and $\Delta\eta$ dependence of the second moments of net-protons has been measured and a deviation from the Skellam baseline is observed.  However, a model including the effects of volume fluctuations and baryon number conservation is able to fully describe the difference without needing additional fluctuations, therefore indicating that these measurements agree with LQCD predictions.  

The cross-species correlated fluctuations of pions, kaons, and protons are also measured with the observable $\nu_{dyn}$ and show qualitative, but not quantitative, agreement with Monte Carlo models.  While there is a sign change observed in $\nu_{dyn}[p,K]$ and $\nu_{dyn}[\pi,p]$ from RHIC to LHC energies, the energy dependence shows a smooth behavior over the large range.  Future investigations and measurements of the higher moments in ALICE will more fully explore the physics of the fluctuations of conserved charges in heavy-ion collisions.  
\\
\\
	
\end{document}